# A Discrete Packing Model of Granular Material Confined in a Vertical Column

Qinghai Jiang[1]; Kai Wu[1]*; Yu Sun[1]; Xin Xie[2]; Zhengyu Yang[2]

1. School of Mechanical Engineering, Nanjing University of Science and Technology, Nanjing, China.

2. Department of Electrical and Computer Engineering, Northeastern University, Boston, MA 02115

   *wukai@njust.edu.cn

**Abstract:** In this paper, we analyzed the transmission rules of interparticle forces between granular particles, based on which, we then proposed a discrete packing model to calculate the static pressure at the bottom of granular material confined in a vertical column. Our mechanical analysis and numerical simulation results indicate that the silo effect is caused by the frictional contacts between border particles and inner walls, the static pressure at the bottom depends on the external load initially, and then tends to a saturation pressure ($P_n$) in an exponential form. The saturation pressure is positive linear related to the container radius ($R$) with the same granular matter and stacking manner. The saturation pressure is directly proportional to the particle size ($r_a$), and the increasing or decreasing characteristic depends on the frictional property of inner walls, the friction and stacking angle of grains. Finally, we compared the predictions of the aforementioned model with the experimental results from the literature, and we observed that good agreement is achieved.

## 1 Introduction

The mechanical status of granular matter is presently one of the most open and hotly debated issues. It is estimated that the transportation and stockpiling energy consumption of granular matter accounts for more than 10% of total energy in the global. Consequently, the investigation of granular matter will be significantly important for reducing energy consumption. The state of granular matter exhibits many unusual mechanical properties and pretty unique phenomena, such as the silo effect (Janssen, 1895) and the ratchet effect, due to the complex force transmission and distribution among particles. The pioneering work on silo effect was carried out by Janssen. Although Janssen model is generally accepted and applied, some experimental studies indicate that the redirection parameter *J* in Janssen theory is not a constant but affected by the ratios of diameter of silo to particle (Qadir *et al.*, 2016) and stacking angle (Lin *et al.*, 2015).

Because of the lack of discrete parameters in Janssen model, some experimental and numerical researches aiming at the dependence of apparent mass at the bottom of grain columns on grain size, container diameter and packing fraction have been carried out. Physics simulation is a feasible way to investigate the complex mechanical process (Jin *et al.*, 2015). However, few theoretical models are investigated where the variations of the apparent mass with these discrete characters in a granular column are involved. In this study, a discrete packing model of granular material with slightly different dimensions confined in a vertical column was proposed and analyzed. The quasi-static process is considered without time effects and only the statics of silos will be discussed.

## 2 The transmission rules of interparticle forces

Aiming at force distribution of dry and cohesionless granular matter confined in a vertical column, some assumptions are proposed to simplify some particle features as following.
1   Particles are cohesionless and treated as planar and rigid discs in two-dimensional.
2   The friction between granular particles and inner wall is fully mobilized.

According to the position and force state, the particles can be divided into border particles and internal particles, and the force states are shown in Fig.1, respectively.

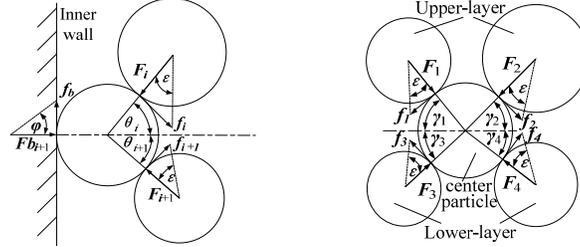

Fig. 1 Force status of border particles and internal particles

The particle is temporarily assumed to be smooth, frictionless and weightless. The transfer mechanism of interparticle forces can be calculated by the force balance

For border particles, the interparticle forces are transmitted from upper-layer to the lower-layer through centre particles according to the rule shown in equation (1)

$$\left.\begin{aligned} F_{i+1} &= F_i \sin\theta_i / \sin\theta_{i+1}; \\ Fb_{i+1} &= F_i \sin(\theta_i + \theta_{i+1}) / \sin\theta_{i+1} \end{aligned}\right\} \quad (1)$$

where: $F_{i+1}$ and $F_i$ are the contact forces of border particle with neighbor particles, $Fb_{i+1}$ is the friction between border particle and silo walls, $\theta_i$ and $\theta_{i+1}$ are the stacking angles.

For internal particles, the interparticle forces are divided into major and minor parts, and transmitted through centre particle from upper-layer to lower-layer in the form of cross-transfer according to the law as shown in equation (2)

$$\left.\begin{aligned} F_3 &= \alpha F_2 + \beta F_1; \\ F_4 &= \alpha' F_1 + \beta' F_2 \end{aligned}\right\} \quad (2)$$

where:

$$\left.\begin{aligned} \alpha &= \sin(\gamma_2 + \gamma_4)/\sin(\gamma_3+\gamma_4); \quad \beta = \sin(\gamma_1-\gamma_4)/\sin(\gamma_3+\gamma_4) \\ \alpha' &= \sin(\gamma_1+\gamma_3)/\sin(\gamma_3+\gamma_4); \quad \beta' = \sin(\gamma_2-\gamma_3)/\sin(\gamma_3+\gamma_4) \end{aligned}\right\} \quad (3)$$

Since $\alpha > \beta'$ and $\alpha' > \beta$, the major part of interparticle forces is transmitted from upper-layer particle to the diagonal particle in lower-layer, and correspondingly the minor part to the particle below. It means that the transmission of strong force exhibits directivity along the lineation of particle centres in contact roughly, basically in accordance with the conclusion of literature (Sun *et al.*, 2010). In the case of ignoring gravity and friction, the average pressure of particle layer remains constant with increasing of particle layers under external loading. By considering gravity but ignoring friction, the average pressure increases linearly with particle layers. Therefore, frictions are the basic reason of the unique mechanical performance for granular matters. In order to consist with the actual situation of granular matter, the influence of gravity and friction are both reconsidered.

For border particles, the force balance is shown in equations (4) and (5).

$$\left.\begin{array}{l} Fb_{i+1} \cos\varphi = F_i \cos(\theta_i + \varepsilon) + F_{i+1} \cos(\theta_{i+1} + \varepsilon); \\ Fb_{i+1} \sin\varphi + F_{i+1} \sin(\theta_{i+1} + \varepsilon) = F_i \sin(\theta_i + \varepsilon) + M_{i+1} g \end{array}\right\} \quad (4)$$

$$\left.\begin{array}{l} Fb_{i+1} = \left(F_i \sin(\theta_i + \varepsilon + \theta_{i+1} + \varepsilon) + M_{i+1} g \cos(\theta_{i+1} + \varepsilon)\right) / \left(\sin(\theta_{i+1} + \varepsilon + \varphi)\right) \\ F_{i+1} = \left(F_i \sin(\theta_i + \varepsilon - \varphi) + M_{i+1} g \cos\varphi\right) / \left(\sin(\theta_{i+1} + \varepsilon + \varphi)\right) \end{array}\right\} \quad (5)$$

where: $\varepsilon$ is the internal friction angle of granular particles, $\varphi$ is the friction angle between granular particles and inner wall.

It can be seen that the interparticle force is weakened according to a specified proportion and transmitted to the lower-particles, and the transfer direction of which is reflected by the container wall.

For internal particles, the force balance is shown in equations (6) and (7).

$$\left.\begin{array}{l} F_1 \cos(\gamma_1 + \varepsilon) - F_2 \cos(\gamma_2 + \varepsilon) = F_4 \cos(\gamma_4 + \varepsilon) - F_3 \cos(\gamma_3 + \varepsilon); \\ F_1 \sin(\gamma_1 + \varepsilon) + F_2 \sin(\gamma_2 + \varepsilon) + M_{i+1} g = F_3 \sin(\gamma_3 + \varepsilon) + F_4 \sin(\gamma_4 + \varepsilon) \end{array}\right\} \quad (6)$$

$$\left.\begin{array}{l} F_3 = \dfrac{\sin(\gamma_2 + \gamma_4 + 2\varepsilon)}{\sin(\gamma_3 + \gamma_4 + 2\varepsilon)} F_2 + \dfrac{\sin(\gamma_1 - \gamma_4)}{\sin(\gamma_3 + \gamma_4 + 2\varepsilon)} F_1 + \dfrac{M_{i+1} g \cos(\gamma_4 + \varepsilon)}{\sin(\gamma_3 + \gamma_4 + 2\varepsilon)}; \\ F_4 = \dfrac{\sin(\gamma_1 + \gamma_3 + 2\varepsilon)}{\sin(\gamma_3 + \gamma_4 + 2\varepsilon)} F_1 + \dfrac{\sin(\gamma_2 - \gamma_3)}{\sin(\gamma_3 + \gamma_4 + 2\varepsilon)} F_2 + \dfrac{M_{i+1} g \cos(\gamma_3 + \varepsilon)}{\sin(\gamma_3 + \gamma_4 + 2\varepsilon)} \end{array}\right\} \quad (7)$$

Comparing equation (2) with equation (7), the transfer mechanism of interparticle forces under gravity keeps consistency with that without gravity, approximately. Therefore, it can be concluded that the silo effect is caused by the friction of border particles rather than internal particles.

Because measuring interparticle forces directly is difficulty in microcosmic level, silo effect is commonly investigated through measurement of the macroscopic parameters such as wall friction (Bek *et al.*, 2016) and base pressure (Qadir *et al.*, 2016). The pressure is consisted of vertical component of interparticle forces in the particle layer. The horizontal and vertical components can be calculated as equation (8).

$$\left.\begin{array}{l} Fx_i = F_i \cos(\theta_i + \varepsilon); \\ Fy_i = F_i \sin(\theta_i + \varepsilon) \end{array}\right\} \quad (8)$$

Equation (8) is substituted into equation (4), then we get

$$\left.\begin{array}{l} Fb_{i+1} \cos\varphi = Fx_i + Fx_{i+1} = Fy_i / \tan(\theta_i + \varepsilon) + Fy_{i+1} / \tan(\theta_{i+1} + \varepsilon); \\ Fb_{i+1} \sin\varphi + Fy_{i+1} = Fy_i + M_{i+1} g \end{array}\right\} \quad (9)$$

through simplification, it reduces to

$$Fy_{i+1} = p_i q_i Fy_i + q_i M_{i+1} g \qquad (10)$$

where, $p_i$ and $q_i$ are two constants related to the stocking manner, the surface roughness of particles and inner wall, and can be calculated by equation (11)

$$p_i = 1 - tan\varphi / tan(\theta_i + \varepsilon); q_i = 1 - tan\varphi / (tan(\theta_{i+1} + \varepsilon) + tan\varphi) \qquad (11)$$

For border particles, the transference pressure is composed of two components: pressure transfer and gravity effect. Obviously, $p_i<1$ and $q_i<1$. It means that the vertical component of interparticle force is transmitted to the lower-layer with an attenuation coefficient of $p_i q_i$, while the gravity component with an attenuation coefficient of $q_i$.

## 3 Analysis and Discussion

### 3.1 Parameter analysis of the two-dimensional model

Based on the transmission rules proposed above, the force network and pressure distribution of granular matter confined in a two-dimensional column can be calculated. To simplify the calculation, particles are aligned in particular arrangement layer by layer, all the particles in the same layer have the same diameter. The particles layers are numbered 1, 2… $n$ from top to bottom, with the radius of $r_1$ to $r_n$, respectively. The contact points of particles are numbered 1, 2 … $2m$ from left to right, with the contact force of $f_1$ to $f_{2m}$. Assuming that $\theta_i$ is the stocking angle between lineation of two neighbor particle centres in different layers and horizontal plane, and $r_m$ is half of the distance between two neighbor particle centres in the same layer. The arrangement of particle and interparticle forces in a vertical column is shown in Fig.2.

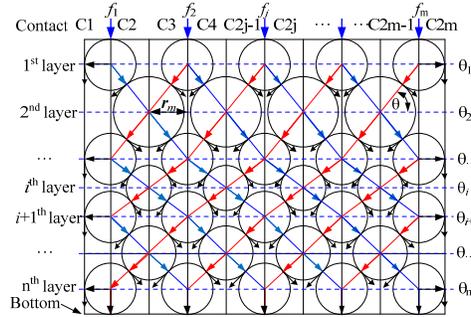

Fig. 2 The force transfer between piles with different diameter

As a numerical example, the initial states can be obtained with feature parameters as following: $r_m = 7.5$mm, $f = 0$, $m = 10$, $\varphi = 15°$, $\varepsilon = 15°$. Figure 3 depicts the dependence of the interparticle forces scaled with the maximum value of interparticle forces respectively on the particle layers under different external load in a 2-D silo.

It can be observed that for zero external load, the interparticle forces are very small and increase slowly with particle layers. The force distributions are roughly uniform in the upper part. In the middle part, some interparticle forces increase with particle layers and form strong force chains. In the lower part, the number of strong force chains increase significantly, and most of strong force chains tend to be similar. In the condition of large external load on the top uniformly, the interparticle forces are much bigger than that of no external load, and two large force chains from the top edge to the opposite wall below can be observed in the upper part. In the middle part, most of the force chains are weakened significantly. In the lower part, the force chains are further weakened and tend to be similar. For the status of point load, it can be observed that the strong force chains originated from the point load are conspicuous. With the increasing of particle layers, the force distribution tends to be uniform basically.

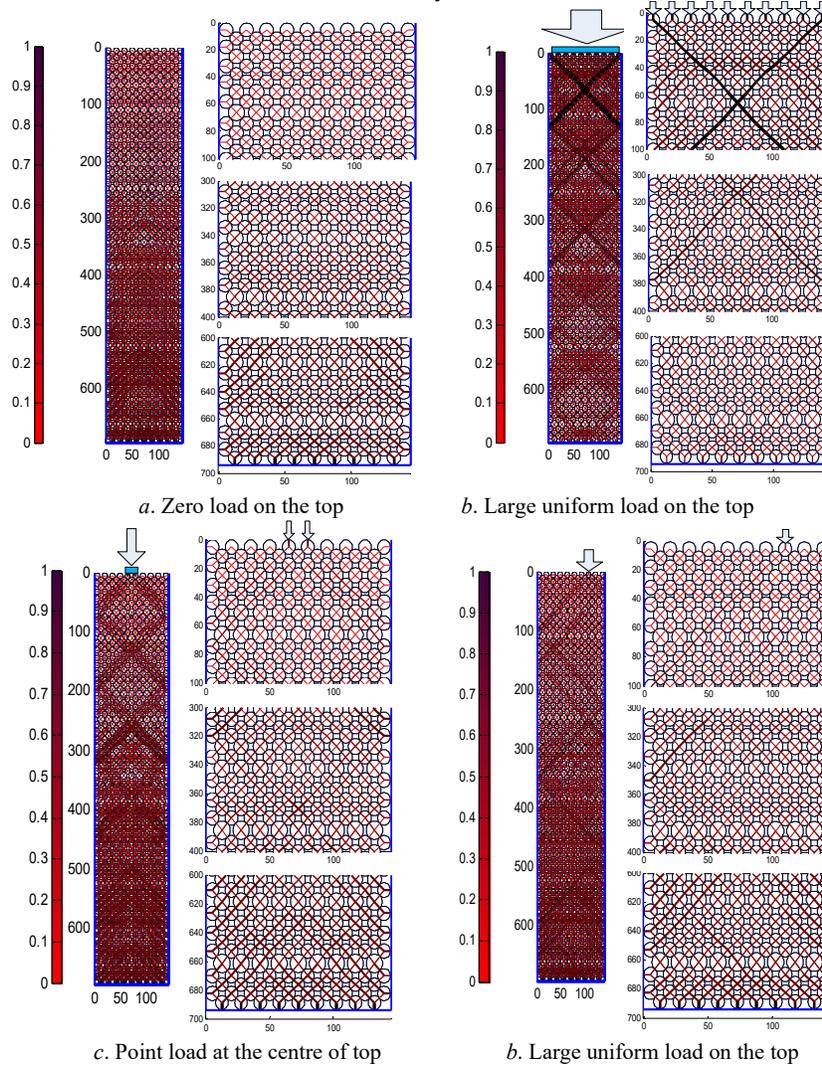

*a*. Zero load on the top  *b*. Large uniform load on the top

*c*. Point load at the centre of top  *b*. Large uniform load on the top

Fig. 3 The interparticle forces distribution in a vertical 2-D column

According to the force distributions under different condition, some common characteristics can be found, the strong force chains are approximately in the same direction toward to the container walls below, rather than to the bottom of column directly. Force chains extend to the container walls below with growing up, and then change their transmission direction with weakened in a certain degree at the inner walls.

Those large force chains, which are significantly greater than others, will be reduced more. As a result, most of the force chains tend to be approximately uniform.

Based on the recursive relations mentioned above, the base pressure curves under external load uniformly can be obtained by MATLAB, as shown in Fig.4. It is shown that the base pressure curves are determined by external load initially and then tend to the saturation pressure ($P_n$) in an exponential form, and fluctuate around a certain level.

For rising curve, gravity plays a major role under zero or small external load with a few particles. At the beginning, base pressure is basically linear proportion to particle layer, as the attenuation of base pressure can be neglected compared to gravity. With the increasing of particle layers, base pressure and attenuation increase progressively. Thus, the base pressure tends to saturation pressure with the decreasing of difference between pressure attenuation and gravity. Silo effect is a typical representative of this situation. For decreasing curve, base pressure plays a major role under an external load much larger than gravity with a few particles, and the influence of gravity can be neglected basically. With the increasing of particle layers, the base pressure decreases rapidly. And the base pressure tends to be stable around the saturation pressure ($P_n$) with the attenuation of base pressure approaching to gravity that can't be ignored gradually. Uniaxial compression of granule is a typical representative of this situation.

Based on the recursive relations mentioned above, the base pressure curves with external load $f = 100$ can be obtained at m=10, 15 and 20, as shown in Fig.5.

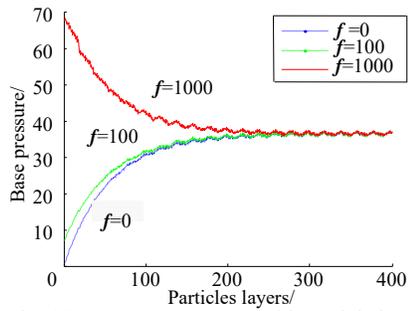 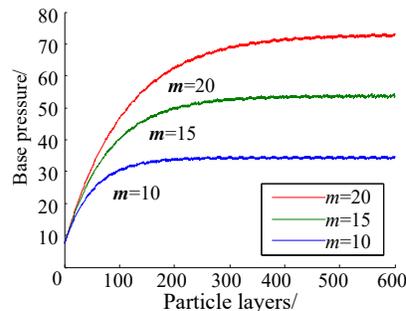

Fig. 4 Base pressure curves with particle layers under different external load

Fig. 5 Base pressure curves with particle layers at different container diameter

Under the same boundary condition, attenuation coefficient depending on the value of $p$ and $q$, are the same for each individual particle. But the attenuation coefficient for a particle layer decreases with the increasing of the diameter of container, because the border rate ($\eta$) in a horizontal plane decreases. Border rate is defined as the ratio of number of border particles to total particles ($N_b/N_T$) in a particle layer and determined by many parameters such as stacking manner and container and size. As shown in Fig.5, the saturation pressure ($P_n$) changes linearly with the radius ratio basically.

### 3.2 The derivation and discussion of the three-dimensional model

The border particles are generally all around the container wall, in order to make the analysis model closer to the actual condition of granular matter, the transmission rules of interparticle forces are expanded to the three-dimensional condition.

*Title*

According to the preceding analysis, the load on each individual particle is various. However, the vertical load on the *i*th particle layer can be considered as average distribution at the value of $Q_i/N_T$, due to the uniform distribution of granular matter and the cross-transfer of inter particle forces. The average attenuation coefficients ($p$ and $q$) are related to the stacking manner and the surface roughness of container. The border particles and internal particles assume the load of $\eta Q_i$ and $(1-\eta)Q_i$ in the *i*th particle layer, respectively. And the border particles assume the load of $pq\eta Q_i+q\eta M c_i g$, transmitted from neighboring particle layer with attenuation, due to the friction of inner wall. $Mc_i$ is the mass of the whole particle layer. It can be seen from the previous analysis that the vertical load is transmitted between internal particles without attenuation, and the internal particles assume the load of $(1-\eta)Q_i+(1-\eta)Mc_i g$. Therefore, the recursive relation between the $(i+1)$th and $i$th particle layers is shown in equation (12).

$$Q_{i+1} = \left(1-(1-pq)\eta\right)Q_i + \left(1-(1-q)\eta\right)M_{ci}g \qquad (12)$$

And the static load at the bottom can be calculated by equation (13) with the recursive relation.

$$Q_n = \left(1-(1-pq)\eta\right)^{k_nH/r_m}\sum_1^m f + \frac{1-\left(1-(1-pq)\eta\right)^{k_nH/r_m-1}}{(1-pq)\eta}\left(1-(1-q)\eta\right)\frac{\rho Sr_m}{k_n}g \qquad (13)$$

Setting the external load $Q_0=\Sigma f$, and the particle layers $n=k_nH/r_m$, the mass of the whole particle layer $M_{ci}=\rho Sr_m/k_n$, then equation (13) can be written as

$$Q_n = \left(1-(1-pq)\eta\right)^{k_nH/r_m}\sum_1^m f + \frac{1-\left(1-(1-pq)\eta\right)^{k_nH/r_m-1}}{(1-pq)\eta}\left(1-(1-q)\eta\right)\frac{\rho Sr_m}{k_n}g \qquad (14)$$

where: $Q_n$ is the static load at the bottom of container, $k_n$ is a constant related to particle size and stacking manner, $H$ is the filling height of granular matter, $\rho$ is the bulk density of granular matter, $S$ is the sectional area of container.

It can be seen that, with the increasing of particle layers, the external component decreases to zero gradually, and then the static load at the bottom tends to be saturated around the saturation load $Q_{sat}$, as shown in equation (15).

$$Q_{sat} = Q_{n\to\infty} = \frac{\left(1-(1-q)\eta\right)\rho Sr_m g}{(1-pq)\eta k_n} \qquad (15)$$

According to the recursive equation (12), the saturation load $Q_n$ can be obtained.

$$Q_n - Q_{n-1} = M_{ci}g - \left((1-pq)\eta Q_{n-1}+(1-q)\eta M_{ci}g\right) \qquad (16)$$

It can be observed that the fluctuations of saturation load $Q_{sat}$ comes from the deviation between the gravity of the whole added particle layer and the attenuation effect of border particle in the added particle layer. As a result, the smaller the mass of particle layer stacked at the bottom is, the less the fluctuation of saturation load $Q_{sat}$ is. The mass of particle layer increases with the particle diameter in the same container, therefore the fluctuation of saturation load $Q_{sat}$ is related to the particle size, consisting with the conclusion in literature (Chand *et al.*, 2012).

3.3 Model validation

Border rate ($\eta$) can be calculated by the ratio of the number of grains at the circumference to that all over the layer in cylinder silo, which is used most extensively. It can be speculated that the arbitrary neighborhood particles are distributed in an equilateral triangle with one sixth circle in each point in the same cross section, as shown in Fig.6.

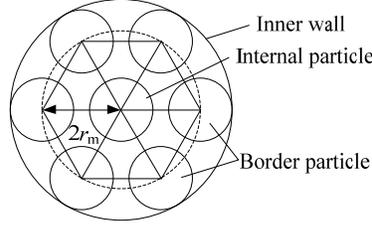

Fig. 6 Schematic diagram of border rate

When container is much larger than particle, the circle formed by the centre of border particles contains all the internal particles and half of the border particles. It can be calculated that the number of border particles is approximately $\pi(R-r_a)/r_m$ and the total particle number in the same layer can be calculated by equation (17) approximately with a geometrical relationship,

$$N_T = \frac{\sqrt{3}\pi(R-r_a)^2}{6r_m^2} + \frac{1}{2}\frac{\pi(R-r_a)}{r_m} \tag{17}$$

where: $N_T$ is the total number of particle in a particle layer, $r_a$ is the mean radius of particles, $r_m$ is the mean nominal radius of particles.

Setting $\delta = (R-r_a)/r_m$ as section filling coefficient, and the border rate ($\eta$) can be expressed by equation (18),

$$\eta = 6/\left(\sqrt{3}\delta + 3\right) \tag{18}$$

where: $\delta$ is a size effect parameter related to the particle size and container diameter. Because the container is generally much larger than particles, the section filling coefficient ($\delta$) can be approximately considered as the ratio of radius of container to nominal radius of particles, $\delta = R/r_m$.

Substitute $\eta$ and $\delta$ into formal (15), then the saturation pressure can be determined by

$$P_n = AR + Br_a \tag{19}$$

where: $A$ and $B$ are two constants related to the stocking manner, the surface roughness of particles and inner wall, and can be calculated by equation (20).

$$A = \frac{\sqrt{3}\rho g}{6k_n(1-pq)}; \quad B = \frac{\rho g k_r}{6k_n(1-pq)}\left(6q - 3 - \frac{\sqrt{3}}{k_r}\right); \quad k_r = r_m/r_a \tag{20}$$

It can be observed that the saturation pressure ($P_n$) depends on many parameters such as stacking manner ($k_n$, $k_r$), container radius ($R$) and particle size ($r_a$). While keeping other conditions invariable, the saturation pressure is linear related to container radius, which is consistent with the experimental result of Liu (Liu *et al.*, 2009), and the increasing rate is related to the density of granular matter ($\rho$), stacking manner ($k_n$) and friction property ($p$, $q$). Correspondingly, the saturation pressure changes linearly with particle size ($r_a$) when other parameters keep constant. Although the model proposed in this paper and Janssen model both embody the linear relation between radius of container and saturation pressure, the Janssen model based on the continuous mechanics theory ignores the effects of particle size and stacking manner. The present model, by contrast, is closer to the real situation by considering the discrete characteristics of granular material.

*Title*

A series of experiments about static load at the bottom were carried out with particle size of 3mm and 5mm in literature (LI *et al.*, 2010). The experimental data and the saturation pressure are shown in Table 1 and two straight lines are fitted with the least square method, as shown in Fig.7.

**Table 1**. Static load and saturation pressure at the bottom with different container and particle sizes

| Container Diameter $D$/ mm | Saturation Load $Q_{3mm}$/ g | Saturation Pressure $P_{3mm}$/pa | Saturation Load $Q_{5mm}$/g | Saturation Pressure $P_{5mm}$/pa |
|---|---|---|---|---|
| 21.3 | 7.46 | 205.17 | 8.04 | 221.12 |
| 27 | 15.35 | 262.73 | 13.29 | 227.48 |
| 33.8 | 35.79 | 390.90 | 26.56 | 290.90 |
| 47.2 | 98.03 | 549.05 | 68.98 | 386.35 |
| 68.2 | 311.82 | 836.51 | 289.01 | 775.32 |
| 103.9 | 1190.26 | 1375.77 | 1078.65 | 1246.77 |

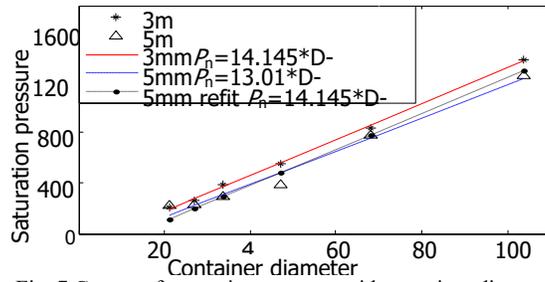

Fig. 7 Curves of saturation pressure with container diameter

The experiment data shows that the increase rates of the two kinds of particles are nearly similar at the same stacking manner. The two pressure lines of 3mm and 5mm particles are almost parallel with correlation coefficient of 0.998 and 0.98 respectively, and in accordance with equation (19) theoretically.

Because the fitting result of 3mm particles is better than 5mm particles, a re-fitting line, parallel to the fitting straight of 3mm particles, is obtained with original data of 5mm particle based on the least square method, as shown in equation (21).

$$P_n = 14.145D - 186 \quad (21)$$

And due to the same material and experimental process, the intercept term in the model is directly proportional to the particle size ($r_a$) in equation (19), and the predicted intercept term of 5mm should be 178.67. Then, the relative error between the predicted and re-fitting value is only 4.1%. It can be observed by the negative constant term that the saturation pressure is reduced with increasing particle size in their experimental condition.

However, a reversed law was observed by Qadir and Jia, the saturation pressure was found augmenting with the grain diameter in their experimental condition. The experimental results of saturation effective masses are 18.1g, 26.3g, 40.8g, 62.4g, 84.6g, 95.0g, 128.0g in the silo with inner diameters of 25.4mm, 29.5mm, 33.1mm, 39.4mm, 43.7mm, 46.3mm respectively (Liu et al., 2009). And the fitting result can be obtained in equation (22) with correlation coefficient of 0.9638 based on the least square method.

$$P_n = 11.017D + 70.559 \quad (22)$$

The positive constant term in Eq. (22) indicates the increasing of saturation pressure ($P_n$) with particle size in the specific experimental conditions. It can be speculated that the two diametrically opposed behaviors seem to be related to the positive and negative characteristics of coefficient **B** in front of *r*<sub>a</sub>, and **B** is finally determined by the friction properties of grains and inner walls, and the stacking angle.

## 4  Conclusions

A discrete model emphasizing the effects of particle characteristics and container radius to the saturation pressure is proposed. A linear relationship between saturation pressure and container radius is found with the same conditions of granular material and stacking manner. It is also found that the saturation pressure changes slightly with the increase of particle size, and the increasing or decreasing characteristic depends on the frictional property of inner walls, the friction and stacking angle of grains. Thereby, the opposite laws between saturation pressure and granular size observed by different scholars can be explained reasonably. The present model has been verified through comparison with existing experimental data with correlation coefficient of 0.998 and 0.98 respectively, and the relative error is only 4.1%.